\documentclass{eas}
\usepackage{graphicx}
\usepackage{epsfig,color,multicol}
\usepackage{astron}

\newcommand{\indx}[1]{ \textrm{\scriptsize #1} }
\newcommand{\bm}[1]{\mbox{{\boldmath $#1$}}}

\begin{document}
\bibliographystyle{astron}
\title{PIERNIK MHD code --- a multi--fluid, non--ideal\\
 extension of the relaxing--TVD scheme~(IV)} 
\runningtitle{M. Hanasz \etal : PIERNIK MHD code \dots ~(IV)}
%
\author{Micha\l{} Hanasz}
\address{Toru\'n Centre for Astronomy, Nicolaus Copernicus University, Toru\'n, Poland; 
\\ \email mhanasz@astri.uni.torun.pl}
\author{Kacper Kowalik}\sameaddress{1} 
\author{Dominik W\'olta\'nski}\sameaddress{1} 
\author{Rafa\l{} Paw\l{}aszek}\sameaddress{1} 
\begin{abstract}
We present a new multi--fluid, grid MHD code PIERNIK, which is based on the
Relaxing TVD scheme~\cite{Jin95}. The original scheme (see Trac \& Pen~\cite*{trac} and Pen~\etal~\cite*{pen}) has been extended by an addition of
dynamically independent, but interacting fluids: dust and a diffusive cosmic ray
gas, described within the fluid approximation, with an option to add other
fluids in an easy way.  The code has been equipped with shearing--box boundary
conditions, and a selfgravity module, Ohmic resistivity module, as well as other
facilities which are useful in astrophysical fluid--dynamical simulations. The
code is parallelized by means of the MPI library.
In this paper we present an extension of PIERNIK, which is  designed for
simulations of diffusive propagation of the Cosmic--Ray (CR) component in the
magnetized ISM.
\end{abstract}
\maketitle
%
%
\section{Cosmic Ray transport}
The CR--MHD extension of PIERNIK code (Hanasz et al. \cite*{hanasz-etal-09a}, 
\cite*{hanasz-etal-09b}, \cite*{hanasz-etal-09c}) is aimed at studies of the
magnetohydrodynamical  dynamo process induced by buoyancy of CRs in stratified
atmospheres of galactic disks (Parker \cite*{parker-92}, Hanasz et al. \cite*{hanasz-etal-04}) investigated previously, in the shearing--box
approximation, with the aid of \ ZEUS-3D code, 
extended with the CR transport algorithm (Hanasz \& Lesch \cite*{hanasz-lesch-03}). 
\par To describe cosmic--ray (CR) propagation in the
interstellar medium (ISM) we use  the diffusion--advection equation (see
Schlickeiser \& Lerche \cite*{schlickeiser-lerche-85}) 
\begin{equation}
\partial_{t} e_{\index{cr}} + \bm{\nabla} (e_{\index{cr}} \bm{v}) 
                  = -p_{\index{cr}} \bm{\nabla}\cdot \bm{v}
             +{\bm{\nabla} (\hat{K} \bm{\nabla} e_{\index{cr}})}+ Q_{\indx{cr}}
	     \label{eqn:crtransport}
\end{equation}
together with the adiabatic equation of state for cosmic rays
\begin{equation}
p_\indx{cr} = (\gamma_\indx{cr} -1) e_\indx{cr},
\end{equation}
in addition to the standard set of MHD equations (Hanasz \& Lesch
\cite*{hanasz-lesch-03}). 
The source term $Q_{\indx{cr}}$ on the rhs. of 
Eqn.~(\ref{eqn:crtransport}) corresponds to the production of CRs in supernova
remnants.  The diffusion term is written in the tensorial form  to account for
anisotropic diffusivity of CRs, where $\hat{K}$ is the diffusion tensor
describing magnetic field--aligned CR diffusion (see Ryu et al
\cite*{ryu-etal-03})
\begin{equation}
K_{ij} = K_{\perp} \delta_{ij} + (K_\parallel - K_{\perp}) n_i n_j, 
\quad n_i = B_i/B,
\end{equation}
\enlargethispage{\baselineskip}
We note that in the presence of CRs an additional source term:  $-\nabla
P_{\indx{cr}}$ should be included in the gas equation of motion (see e.g.
Berezinski et al. \cite*{berezinski-etal-90}), in order to incorporate the
effects of CRs on gas dynamics. 

In order to adopt the CR transport equation to the conservative scheme of
PIERNIK code, we write Eqn.~(\ref{eqn:crtransport}) in the conservative form
\begin{equation}
\partial_t e_{\indx{cr}}+ \bm{\nabla}\cdot \bm{F_{\indx{cr,adv}}}+ \bm{\nabla}
\cdot \bm{F_{\indx{cr,diff}}}
                  = -p_{\indx{cr}} \bm{\nabla} \cdot \bm{v} + Q_{\indx{cr}}, 
\end{equation}
where $e_{\indx{cr}}$ is CR energy density, $\bm{F}_{\indx{cr,adv}} =
e_{\indx{cr}} \bm{v} $, is the flux of CRs advected by the gas flow, 
$\bm{F}_{\indx{cr,diff}} =  -\hat{K} \bm{\nabla} e_{\indx{cr}}$ is CR diffusion
flux and $Q_{\indx{cr}}$  is the CR source term. 
\par The left hand side of the CR transport equation is treated in a conservative
manner, while the terms on r.h.s. are added as source terms. The advection  and
source steps ($p_{\indx{cr}} \nabla \cdot \bm{v}$) for CRs  are implemented within the 
RTVD scheme, while the CR diffusion step is realized outside the Relaxing TVD
routine. The update of CR energy, corresponding to the diffusion term is
performed with the aid of a directionally split, explicit algorithm (Hanasz \& Lesch \cite*{hanasz-lesch-03}), which is first order in time and space. The source step corresponding to the injection of CRs in SN remnants is realized once per double timestep, outside the directional sweeps of fluid updates.
\par The explicit CR diffusion algorithm implemented in PIERNIK code is subject to
the timestep limitation resulting from the von Neumann stability analysis. The timestep
for the diffusive part of the CR diffusion--advection equation imposed in the
code is 
\begin{equation}
  \Delta t = 0.5 \ C_{\indx{cr}} \ 
  \frac{\min (\Delta x,\Delta y, \Delta z)^2}{K_{\parallel} + K_{\perp}},
\end{equation}
where $C_{\indx{cr}} < 1 $ is the Courant number for the CR diffusive transport
algorithm, $\Delta x$,  $\Delta y$ and $\Delta z$ are cell sizes. The numerical
stability of the  overall CRMHD algorithm, is achieved by a proper monotonic
interpolation of CR gradient components computed on cell boundaries (see Hanasz
\& Lesch~\cite*{hanasz-lesch-03}). 
We note, that the appropriate choice of boundary conditions for the highly
diffusive CR component is to set a fixed value (zero) of CR energy density on
external domain boundaries.

\section{Test problems for CR transport}
\begin{figure}[!h]
\centerline{\includegraphics[width=0.37\columnwidth]{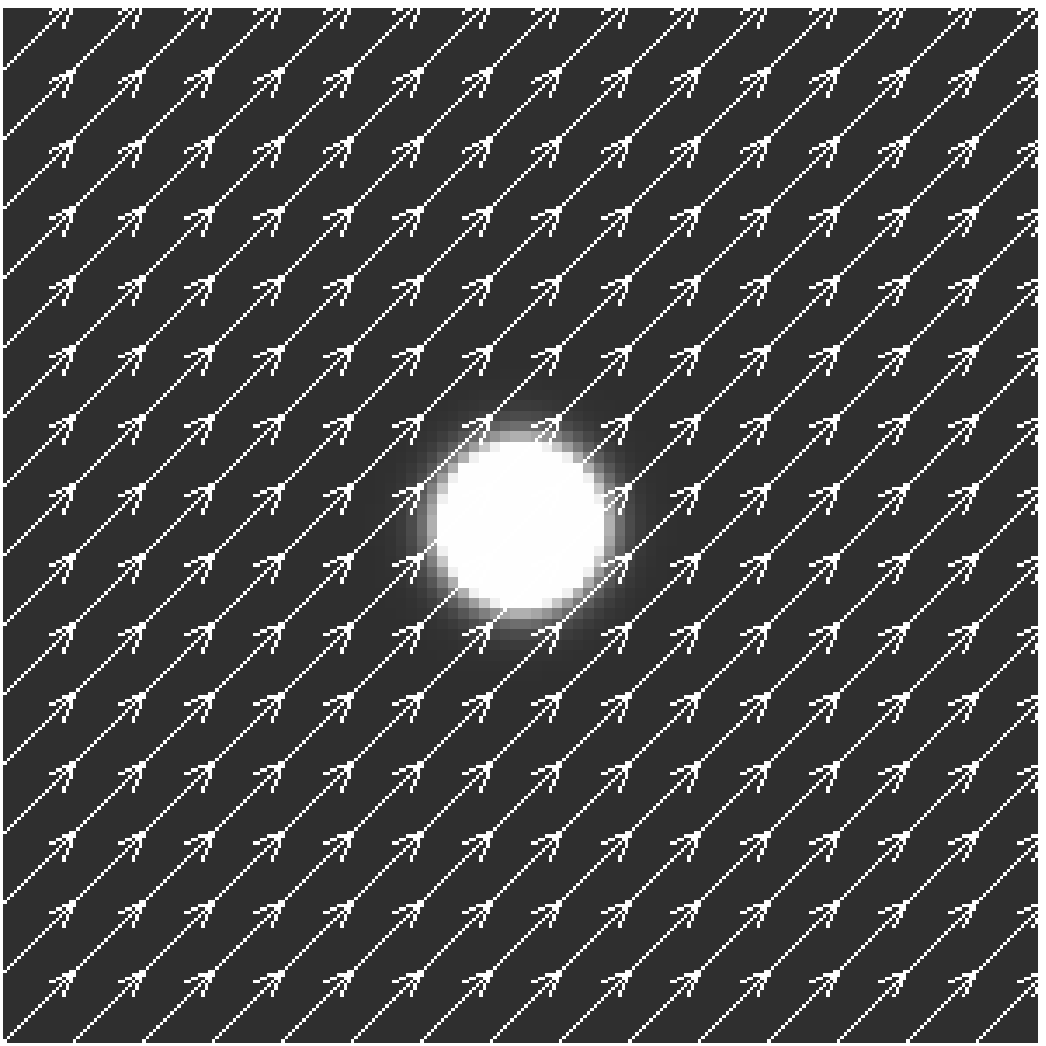}
            \includegraphics[width=0.37\columnwidth]{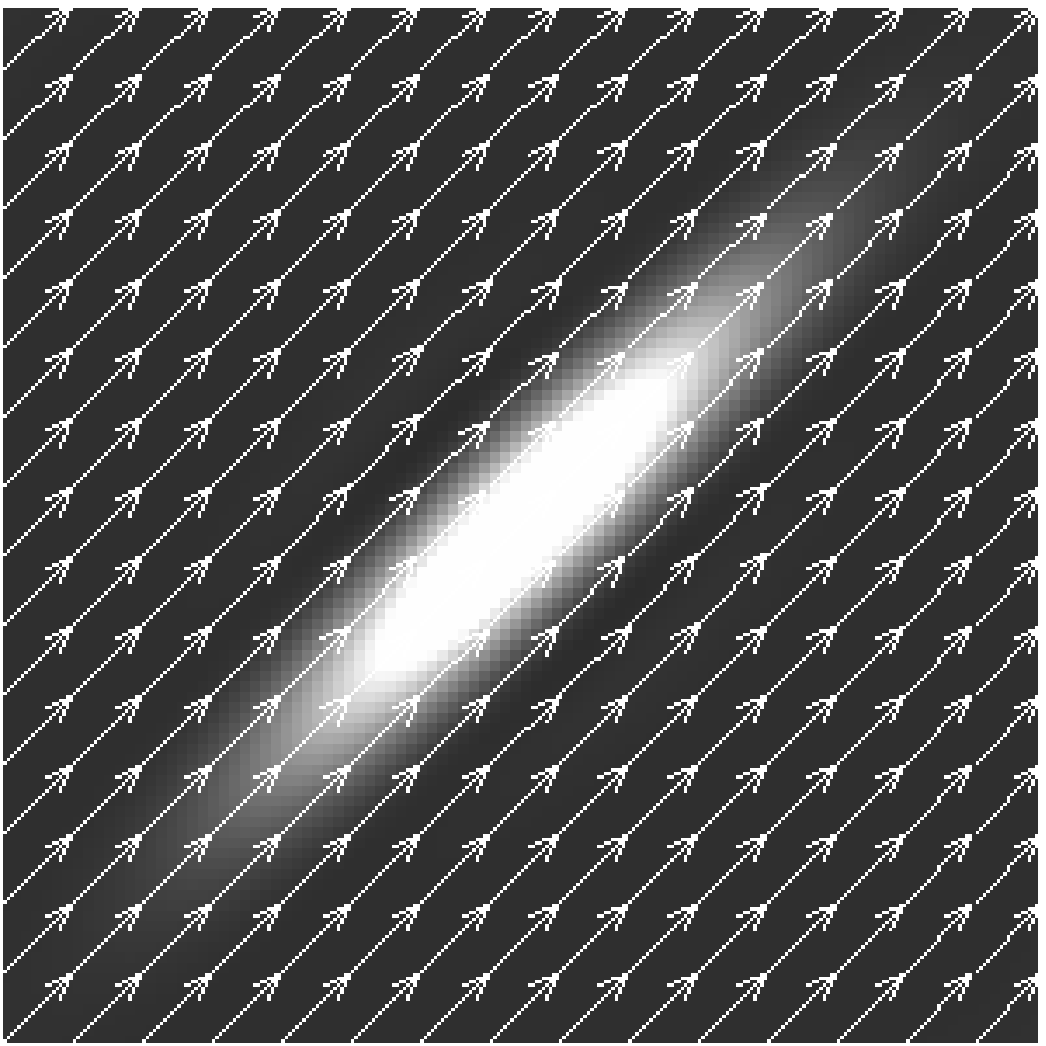}}
\centerline{\includegraphics[width=0.37\columnwidth]{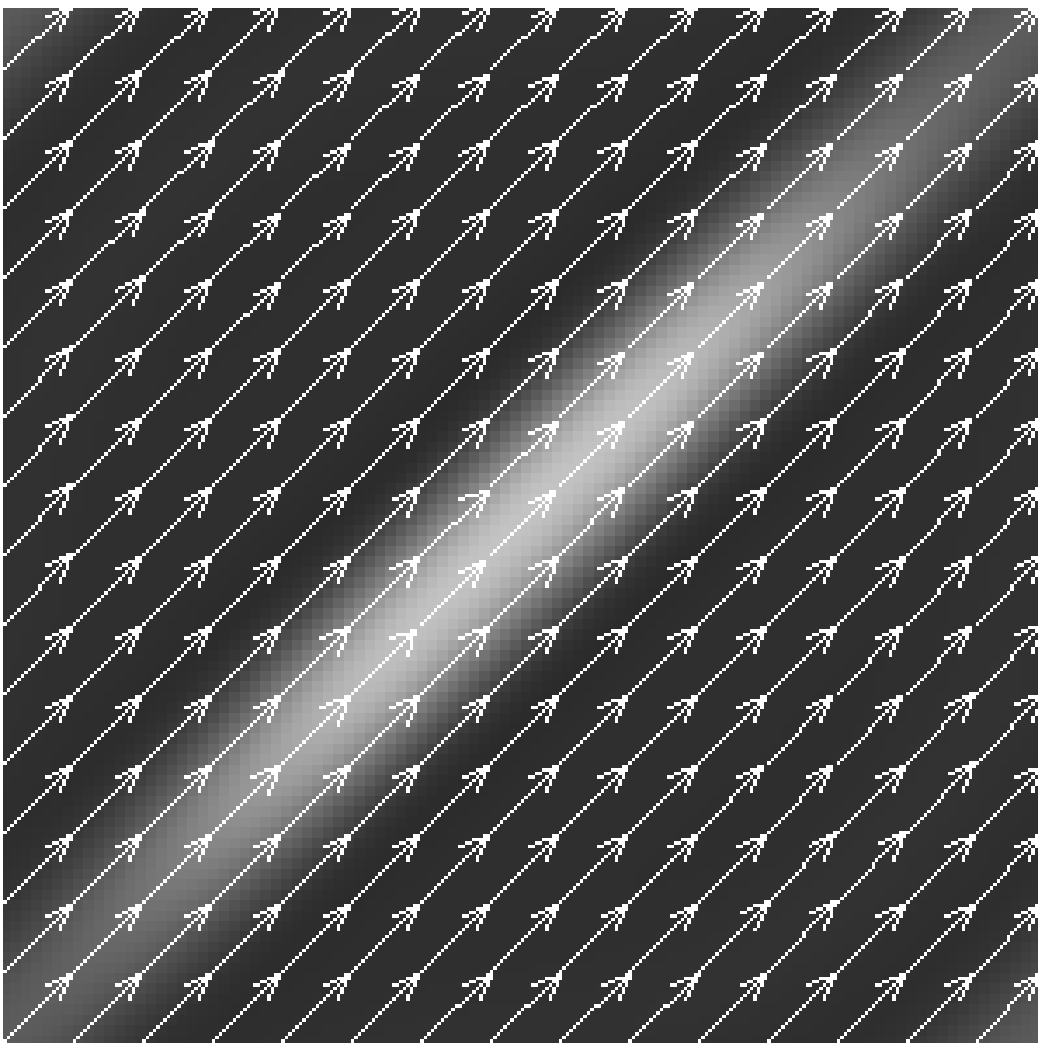}
            \includegraphics[width=0.37\columnwidth]{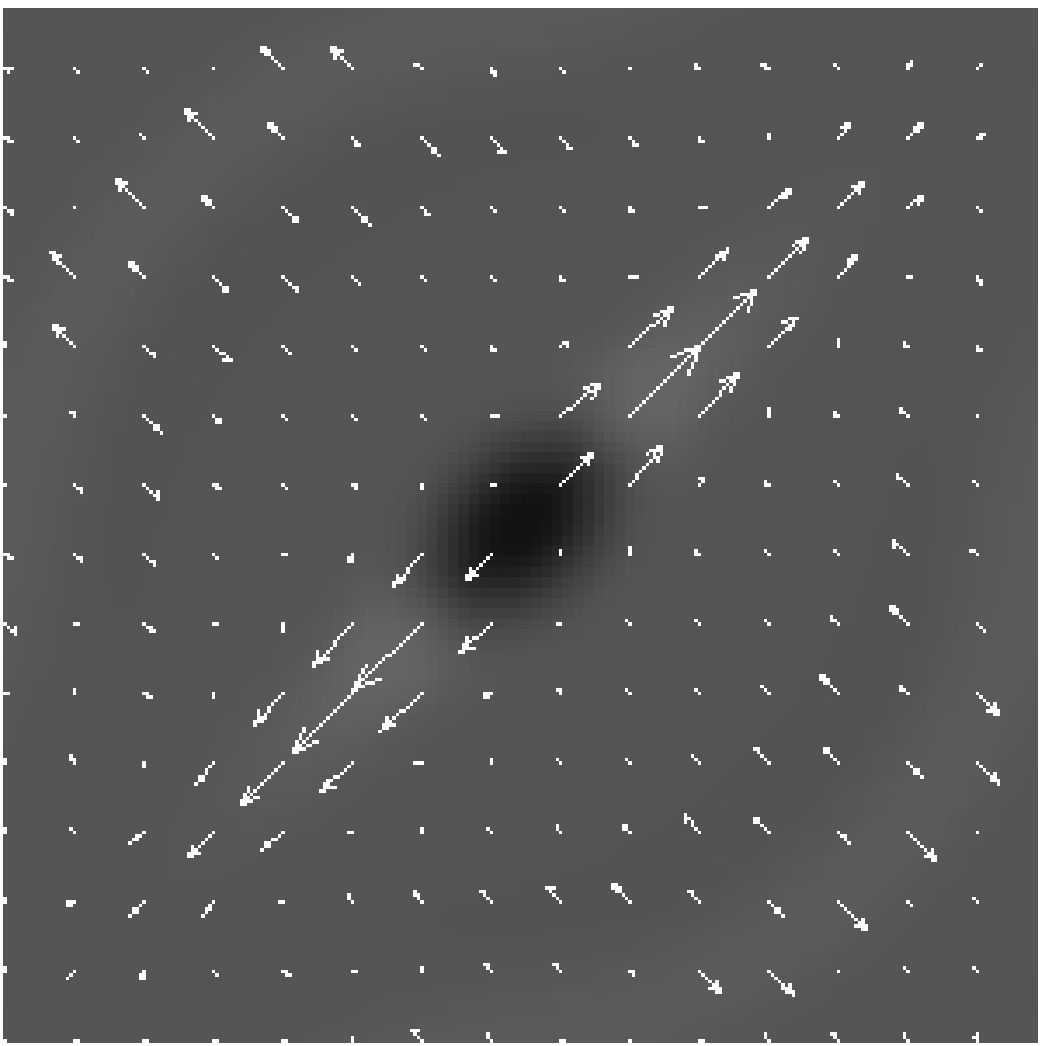}}
\caption{Diffusion of cosmic rays along an inclined magnetic field:  the initial
spheroidal distribution of $e_{\indx{cr}}$ at $t= 0$ and the ellipsoidal
distribution at $t=20$ and $t=60$. The last panel shows thermal gas density and velocity 
vectors at  $t=60$. The apparent flow of gas along the magnetic field direction is due to 
the CR pressure gradient, pushing gas along magnetic field lines. A magnetosonic wave, propagating in
the direction perpendicular to magnetic field is also present.
}
\label{fig:crdiffusion}
\end{figure}
\par To test the magnetic field--aligned cosmic ray diffusion we present a simple 
2D setup with uniform and diagonal magnetic field in the doubly--periodic
computational domain. Parameters of the initial setup for the simulation are (in
arbitrary units): $\rho_0 = 1$, $p_0 = 1$, $B_\indx{x}=3$, $B_\indx{y}=3$, $\gamma = 5/3$, $\gamma_{\indx{cr}}=4/3$,
$x_{\indx{min}}= -500$, $x_{\indx{max}}= 500$, $y_{\indx{min}}= -500$ and $y_{\indx{max}}=
500$. At $t=0$ a portion of CRs, forming a 2D Gaussian profile, with half-width
equal to 50 units  and $e_{\indx{cr}}=8$ at maximum around the domain
center. The diffusion coefficients are $K_\parallel=1000$ and $K_\perp=0$.
\par The results of the test run demonstrate that the CR diffusion proceeds along
magnetic field lines, as expected (see first three panels of
Fig.~\ref{fig:crdiffusion}). A detailed quantitative analysis ensures that
in case of passive CR propagation (without the back-reaction of CR pressure on
the thermal gas) numerical results fit accurately to the analytical solution.
In the present case of active CR propagation, CR pressure gradients affect
thermal gas (see the fourth panel of Fig.~\ref{fig:crdiffusion}).
The excess of cosmic ray pressure near the center of computational domain
accelerates gas, along the oblique magnetic field, forming an ellipsoidal
cavity in the gas distribution. 
\enlargethispage{\baselineskip}
\par The present implementation of CR transport within the very efficient and
flexible Relaxing TVD scheme (Pen et al. \cite*{pen}), combined with the MPI
parallelization of PIERNIK, makes it possible to study the dynamic of CRs, and
CR-driven dynamo in global simulations of galactic disks. First results of global
galactic dynamo simulations (Hanasz et al. \cite*{hanasz-etal-09e},
\cite*{hanasz-etal-09f}) demonstrate that magnetic fields can be efficiently
amplified to equipartition values, on the timescale of galactic rotation, starting
from weak magnetic fields of stellar origin.
\par In a more general case the cosmic ray (CR) component can be considered as an
additional set of fluids extending  the vector $\mathbf{u}$ of conservative
variables. 
A subsequent extension of the CR transport module in PIERNIK code, aiming at energy dependent
treatment of CR--electrons, and incorporation of synchrotron losses is currently
under development.
\section*{Acknowledgements}
\enlargethispage{\baselineskip}
This work was partially supported by Nicolaus Copernicus University through
the grant No. 409--A, Rector's grant No. 516--A, by European Science Foundation within the
ASTROSIM project and by Polish Ministry of Science and Higher Education through
the grants 92/N--ASTROSIM/2008/0 and \mbox{PB 0656/P03D/2004/26}.
%
%

\begin{thebibliography}{}

\bibitem[\protect\astroncite{{Berezinskii} et~al.}{1990}]{berezinski-etal-90}
{Berezinskii}, V.~S., {Bulanov}, S.~V., {Dogiel}, V.~A., and {Ptuskin}, V.~S.:
  1990,
\newblock {\em {Astrophysics of cosmic rays}},
\newblock Amsterdam: North-Holland, ed. by Ginzburg, V.L.

\bibitem[\protect\astroncite{{Hanasz} et~al.}{2004}]{hanasz-etal-04}
{Hanasz}, M., {Kowal}, G., {Otmianowska-Mazur}, K., and {Lesch}, H.: 2004,
\newblock {\em Astrophys. J., Lett.} {\bf 605}, L33

\bibitem[\protect\astroncite{{Hanasz} et~al.}{2009a}]{hanasz-etal-09a}
{Hanasz}, M., {Kowalik}, K., {W{\'o}lta{\'n}ski}, D., and {Paw{\l}aszek}, R.:
  2009a,
\newblock {in K. Go\'zdziewski (eds.), {\em Extrasolar planets in multi--body systems: theory and observations}, (submitted), arXiv:0812.2161}

\bibitem[\protect\astroncite{{Hanasz} et~al.}{2009b}]{hanasz-etal-09b}
{Hanasz}, M., {Kowalik}, K., {W{\'o}lta{\'n}ski}, D., {Paw{\l}aszek}, R., and
  {Kornet}, K.: 2009b,
\newblock {in K. Go\'zdziewski (eds.), {\em Extrasolar planets in multi--body systems: theory and observations}, (submitted), arXiv:0812.2799}

\bibitem[\protect\astroncite{{Hanasz} et~al.}{2009c}]{hanasz-etal-09c}
{Hanasz}, M., {Kowalik}, K., {W{\'o}lta{\'n}ski}, D., and {Paw{\l}aszek}, R.:
  2009c,
\newblock {in M. de Avillez (eds.), {\em The Role of Disk--Halo Interaction in Galaxy
     Evolution: Outflow vs Infall?}, (submitted),  arXiv:0812.4839}

\bibitem[\protect\astroncite{{Hanasz} and {Lesch}}{2003}]{hanasz-lesch-03}
{Hanasz}, M. and {Lesch}, H.: 2003,
\newblock {\em Astron. Astrophys.} {\bf 412}, 331

\bibitem[\protect\astroncite{{Hanasz} et~al.}{2009d}]{hanasz-etal-09e}
{Hanasz}, M., {Otmianowska-Mazur}, K., {Lesch}, H., {Kowal}, G., {Soida}, M.,
  {W{\'o}lta{\'n}ski}, D., {Kowalik}, K., {Paw{\l}aszek}, R., and
  {Kulesza-{\.Z}ydzik}, B.: 2009d,
\newblock {in K.G. Strassmeier, et al. (eds) {\em Cosmic Magnetic Fields: From Planets, to Stars and Galaxies}, Proceedings IAU Symposium No. 259, (submitted), arXiv:0901.0111}

\bibitem[\protect\astroncite{{Hanasz} et~al.}{2009e}]{hanasz-etal-09f}
{Hanasz}, M., {W{\'o}lta{\'n}ski}, D., {Kowalik}, K., and {Paw{\l}aszek}, R.:
  2009e,
\newblock {in K.G. Strassmeier, et al. (eds) {\em Cosmic Magnetic Fields: From Planets, to Stars and Galaxies}, Proceedings IAU Symposium No. 259, (submitted), arXiv:0901.0116}

\bibitem[\protect\astroncite{{Jin} and {Xin}}{1995}]{Jin95}
{Jin}, S. and {Xin}, Z.: 1995,
\newblock {\em Comm. Pure Appl. Math.} {\bf 48}, 235

\bibitem[\protect\astroncite{{Parker}}{1992}]{parker-92}
{Parker}, E.~N.: 1992,
\newblock {\em Astrophys. J.} {\bf 401}, 137

\bibitem[\protect\astroncite{{Pen} et~al.}{2003}]{pen}
{Pen}, U.-L., {Arras}, P., and {Wong}, S.: 2003,
\newblock {\em Astrophys. J., Suppl. Ser.} {\bf 149}, 447

\bibitem[\protect\astroncite{{Ryu} et~al.}{2003}]{ryu-etal-03}
{Ryu}, D., {Kim}, J., {Hong}, S.~S., and {Jones}, T.~W.: 2003,
\newblock {\em Astrophys. J.} {\bf 589}, 338

\bibitem[\protect\astroncite{{Schlickeiser} and
  {Lerche}}{1985}]{schlickeiser-lerche-85}
{Schlickeiser}, R. and {Lerche}, I.: 1985,
\newblock {\em Astron. Astrophys.} {\bf 151}, 151

\bibitem[\protect\astroncite{{Trac} and {Pen}}{2003}]{trac}
{Trac}, H. and {Pen}, U.-L.: 2003,
\newblock {\em Publ. Astron. Soc. Pac.} {\bf 115}, 303

\end{thebibliography}

\end{document}